\documentclass[letterpaper, 10 pt, conference]{ieeeconf}  % Comment this line out if you need a4paper

\usepackage{amssymb}
\usepackage{lscape}
\usepackage{lineno}
\usepackage{graphicx}%
\usepackage{multirow}%
\usepackage{amsmath,amssymb,amsfonts}%

\usepackage{mathrsfs}%
\usepackage{xcolor}%
\usepackage{textcomp}%
\usepackage{manyfoot}%
\usepackage{booktabs}%
\usepackage{algorithm}%
\usepackage{algorithmicx}%
\usepackage{algpseudocode}%
\usepackage{listings}%
\usepackage{tabularx}
\usepackage{array}
\usepackage{booktabs} % For nicer table lines
 \usepackage{multirow}
\usepackage{afterpage,lscape}
\IEEEoverridecommandlockouts                              %
\title{\LARGE \bf
Surg-SegFormer: A 
Dual Transformer-Based Model for Holistic Surgical Scene Segmentation  
}
\author{Fatimaelzahraa Ahmed$^{1}$, Muraam Abdel-Ghani$^{1}$, Muhammad Arsalan$^{2}$, Mahmoud Ali$^{2}$,\\ 
Abdulaziz Al-Ali$^{2}$, and Shidin Balakrishnan$^{1,*}$% <-this % stops a space
\thanks{*Corresponding author.}% <-this % stops a space
\thanks{$^{1}$Department of Surgery, Hamad Medical Corporation, Doha, P.O Box 3050, Qatar. 
        {\tt\small \{fatmaahmed.hmc, sbalakrishnan1\}@hamad.qa}; \tt\small muraam.abdelghani@outlook.com}%
\thanks{$^{2}$College of Engineering, Qatar University, Doha, P.O Box 2713, Qatar.
        {\tt\small \{Muhammad.Arsalan, ma2103840, a.alali\}@qu.edu.qa}}%
}

\begin{document}

\maketitle
\thispagestyle{empty}
\pagestyle{empty}

%%%%%%%%%%%%%%%%%%%%%%%%%%%%%%%%%%%%%%%%%%%%%%%%%%%%%%%%%%%%%%%%%%%%%%%%%%%%%%%%
\begin{abstract}
Holistic surgical scene segmentation in robot-assisted surgery (RAS) enables surgical residents to identify various anatomical tissues, articulated tools, and critical structures, such as veins and vessels. Given the firm intraoperative time constraints, it is challenging for surgeons to provide detailed real-time explanations of the operative field for trainees. This challenge is compounded by the scarcity of expert surgeons relative to trainees, making the unambiguous delineation of go- and no-go zones inconvenient. Therefore, high-performance semantic segmentation models offer a solution by providing clear postoperative analyses of surgical procedures. However, recent advanced segmentation models rely on user-generated prompts, rendering them impractical for lengthy surgical videos that commonly exceed an hour. To address this challenge, we introduce Surg-SegFormer, a novel prompt-free model that outperforms current state-of-the-art techniques. Surg-SegFormer attained a mean Intersection over Union (mIoU) of 0.80 on the EndoVis2018 dataset and 0.54 on the EndoVis2017 dataset. By providing robust and automated surgical scene comprehension, this model significantly reduces the tutoring burden on expert surgeons, empowering residents to independently and effectively understand complex surgical environments.
\end{abstract}

%%%%%%%%%%%%%%%%%%%%%%%%%%%%%%%%%%%%%%%%%%%%%%%%%%%%%%%%%%%%%%%%%%%%%%%%%%%%%%%%

\section{Introduction }

Accurate decision-making in robot-assisted surgery (RAS) requires a thorough understanding using computer vision models \cite{Accurate}. These models need to identify and segment anatomical structures and articulated tools to interpret and understand the relation between objects within the scene \cite{DeepLearningSys}. Nevertheless, accurate and comprehensive surgical scene segmentation remains a significant challenge due to the complexity of anatomical structures and the dynamic nature of the surgical environment. 

Entry-level surgeons can benefit from using these models to convert surgical scenes into self-explanatory videos, as they are not accustomed to how these structures appear in live surgery settings \cite{Entry-level}. Simply, the output video highlights critical zones and detects various articulated tools within the frame. Moreover, such automation frees expert surgeons from suspending the operation to answer the trainees’ questions \cite{Challenge}. Once objects within the surgical scene are accurately identified, understanding the procedure becomes significantly easier.

Cutting-edge segmentation models, e.g., AdaptiveSAM \cite{AdaptiveSAM}, exhibit excellent performance; however, their dependence on manual prompts restricts their autonomy and scalability in real surgical practice. This limitation is particularly significant in post-operative analysis, as surgical videos often exceed three hours, making manual prompting infeasible. In comparison, models like ISINet \cite{ISINet}, SegNet \cite{segnet}, and Ternaus \cite{ternaus} are more efficient and better suited for large-scale automated surgical analysis. Despite their autonomy, the promptable model still outperforms.

To overcome these limitations, we extend SegFormer \cite{SegFormer} by developing a dual-instance pipeline. The first instance employs the SegFormer B2 variant, fine-tuned exclusively for anatomical structure segmentation—referred to as SegAnatomy. The second instance uses B5 variant encoder and incorporates a custom-designed, lightweight decoder optimized for segmenting articulated surgical tools, which we refer to as SegTool. In the end, the outputs of the two instances are fused using a priority-weighted conditional fusion strategy, offering comprehensive and consistent segmentation of surgical frames. We call this complete pipeline Surg-SegFormer.

This paper has three main contributions:
\begin{enumerate}
\item \textit{Dual-Model Segmentation Framework:} A framework for robotic-assisted surgery (RAS) that uses two distinct models specialized in segmenting anatomical structures and surgical instruments.
\item \textit{Priority-Weighted Conditional Fusion Strategy:} An advanced fusion strategy that combines both model outputs, prioritizing valuable segmentation cues to enhance overall accuracy and robustness.
\item \textit{Comprehensive Evaluation on Benchmark Datasets:} Validation of our framework on two benchmark datasets, demonstrating superior segmentation performance compared to current state-of-the-art (SOTA) methods.
\end{enumerate}

\section{Related Work}

Semantic segmentation has long been a core problem in computer vision and is particularly critical in medical‐image analysis and robotic‐assisted surgery (RAS). The introduction of U‐Net \cite{Unet} marked a pivotal milestone: its encoder–decoder design with skip connections quickly became the de facto backbone for biomedical‐image segmentation and inspired a family of surgical‐specific models.

\subsection{CNN-based Approaches}
Subsequent convolutional neural network (CNN) architectures—Mask R-CNN \cite{maskrcnn}, DeepLabV3+ \cite{deeplab}, SegNet \cite{segnet}, and TernausNet \cite{ternaus}, among others—pushed performance further by emphasizing hierarchical feature extraction and multiscale context aggregation. CNNs excel at detecting and segmenting rigid structures under controlled conditions; their receptive fields remain intrinsically local. Consequently, they struggle to capture long-range dependencies and the global context that characterize complex surgical scenes containing occluded, deformable anatomy and tools.

\subsection{Transformer-Based and Hybrid Models}
Recent work has shifted toward transformer-based designs, which employ self-attention to model global relationships. Vision transformers have demonstrated remarkable improvements in image segmentation and pervade the medical domain. SegFormer \cite{SegFormer}, in particular, pairs a lightweight hierarchical transformer encoder with efficient multilayer‐perceptron (MLP) decoders, achieving SOTA performance. Its strong performance across diverse datasets makes it an ideal foundation for downstream tasks such as surgical‐scene understanding.

\subsection{Promptable vs.~~Prompt‑free Paradigms}
Parallel to architectural advances, interaction paradigms for segmentation have diverged into two camps:
\begin{itemize}
    \item \textbf{Promptable models} (also called interactive or prompt‑based models) require explicit user input—points, bounding boxes, or text—to guide mask generation. The Segment Anything Model (SAM)\cite{SAM} is emblematic, offering impressive zero‑shot capability across open‑world images but depending on human prompts at inference time.
    \item \textbf{Prompt‑free models} operate autonomously once trained, generating masks without external cues. Classical CNNs such as ISINet \cite{ISINet}, SegNet, and TernausNet fall into this category, as do more recent transformer variants that remove the need for manual interaction.
\end{itemize}
Promptable models deliver flexibility yet are ill-suited for lengthy surgical videos—often exceeding three hours—where continuous human prompting is infeasible. In contrast, prompt‑free approaches scale better but historically lag behind promptable transformers in accuracy, particularly for small or highly variable objects.

\section{Methodology}

\subsection{Model Overview}

\begin{figure}[ht!]
\raggedright
    \includegraphics[width=0.50\textwidth]{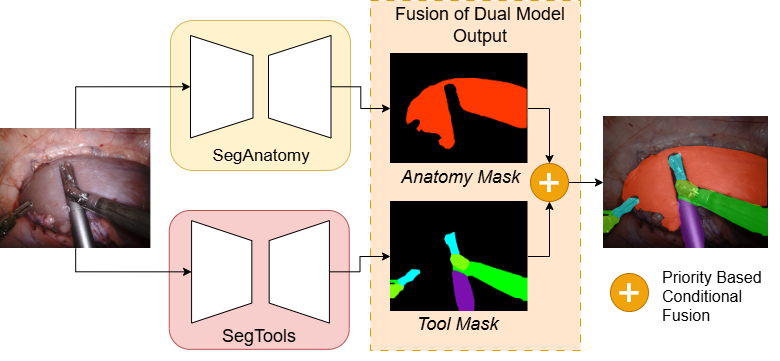}
    \caption{Surg-SegFormer Architecture}
    \label{fig:Surg-SegFormer Architecture}
\end{figure}

We propose \textbf{Surg-SegFormer}, a dual-model framework that leverages two SegFormer instances and fuses their outputs. Figure \ref{fig:Surg-SegFormer Architecture} shows the model’s architecture and the connection between the two instances. The first instance, \textbf{SegAnatomy}, is fine-tuned specifically on anatomical structures. The second instance, \textbf{SegTool}, uses a SegFormer encoder fine-tuned for tool segmentation and replaces the original decoder with a lightweight design that incorporates skip connections. This modification enhances the retention of spatial information—especially for smaller objects like surgical tool tips, which are prone to information loss during down-sampling.
We introduce a priority-weighted conditional fusion strategy to merge the outputs from both instances, ensuring that critical features are preserved in the final segmentation. We evaluate performance using mean intersection over union (IoU) and Dice scores, demonstrating the model’s efficacy in both anatomical and tool segmentation tasks.

\subsection{Surg-SegFormer Architecture}

SegFormer is a hierarchical, transformer-based model that incorporates lightweight multilayer perceptrons (MLPs) and is well suited for semantic segmentation \cite{SegFormer}. The model follows an encoder–decoder architecture: the encoder captures multi-scale features, while the decoder reconstructs high-resolution segmentation masks.

\textbf{\textit{SegAnatomy}} is a fine-tuned adaptation of the SegFormer B2 architecture, selected for its superior performance in segmenting anatomical structures. Table \ref{tab:Variants Comparison} shows that, among the six versions tested, SegFormer B2 achieved the highest mIoU and Dice scores.
\begin{table}[ht!]
\caption{Comparison of mIoU and Dice for Instruments and Anatomy across SegFormer variants}
\label{tab:Variants Comparison}
\centering % IEEE recommends this instead of \begin{center}
\scalebox{1.2}{ % Change the scaling factor to make the table smaller or larger
\begin{tabular}{@{}lllll@{}}
\toprule
\multirow{2}{*}{Variant} & \multicolumn{2}{l}{Instruments} & \multicolumn{2}{l}{Anatomy} \\ \cmidrule(l){2-5} 
 & mIoU & Dice & mIoU & Dice \\ \midrule
B0 & 63.05 & 64.14 & 55.72 & 61.19 \\ \midrule
B2 & 63.80 & 66.32 & \textbf{66.56} & \textbf{74.17} \\ \midrule
B5 & \textbf{75.20} & \textbf{78.34} & 60.76 & 61.72 \\ \bottomrule
\end{tabular}
}
\end{table}

\begin{figure}[ht!]
\raggedright
    \includegraphics[width=\linewidth]{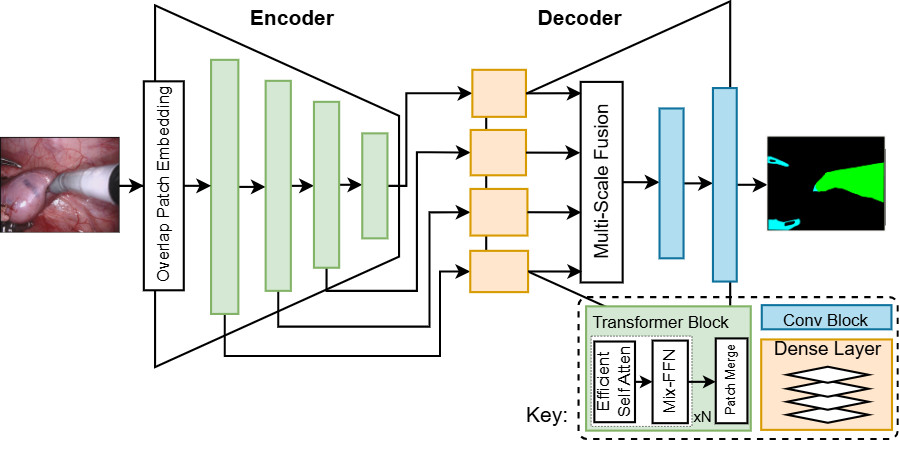}
    \caption{SegTools Architecture.}
    \label{fig:SegTools}
\end{figure}

\textbf{\textit{SegTools}} is a SegFormer–based architecture tailored for precise tool segmentation. We selected the SegFormer-B5 variant because it achieved the best results among the variants tested (Table \ref{tab:Variants Comparison}). To better retain spatial information for classes with limited pixel representation, we utilized the original encoder and designed a custom decoder. This change addresses a limitation in the standard SegFormer encoder–decoder architecture: the input image undergoes progressive downsampling across four stages—by factors of 4, 8, 16, and 32—resulting in greatly diminished spatial detail being passed to the decoder. Reconstructing fine-grained structures, such as surgical tools, thus becomes challenging, as they may no longer be discernible at that scale. Each stage involves patch embedding and merging, which reduces spatial dimensions while increasing channel depth. For an input image \(\mathbf{I} \in \mathbb{R}^{H \times W \times 3}\), the feature map at stage \(i\) is represented as:
\[
\mathbf{X}^{(i)} \in \mathbb{R}^{B \times C_i \times \frac{H}{2^i} \times \frac{W}{2^i}}
\]
To overcome this, our lightweight decoder connects densely to the encoder layers via skip connections (Figure \ref{fig:SegTools}), preserving fine-grained details by integrating multi-scale information. The custom decoder comprises three key components:
\begin{enumerate}
    \item \textbf{Uniform Channel Projection and Upsampling:} Each encoder feature map \(\mathbf{X}^{(i)}\) is projected to a uniform channel dimension using a linear projection (\(\mathbf{x}_i = \mathrm{Proj}_i(\mathbf{X}^{(i)})\)) and then upsampled to the highest resolution.
    \item \textbf{Multi-Scale Feature Fusion:} The upsampled features \(\mathbf{x}_1^{\mathrm{up}}\), \(\mathbf{x}_2^{\mathrm{up}}\), \(\mathbf{x}_3^{\mathrm{up}}\), and the highest-resolution feature \(\mathbf{x}_4\) are concatenated into a single tensor, preserving information across scales for accurate segmentation of objects of varying sizes.
    \item \textbf{Dense Skip Connections:} These connections revisit fine details from earlier stages throughout the decoding process, enhancing the model’s ability to focus on small, intricate structures.
\end{enumerate}
This customized architecture effectively mitigates the challenges posed by extensive downsampling, improving the model’s capability to accurately segment tools and other small structures in medical images.

\textbf{\textit{Final segmentation}} output is derived by fusing predictions from both instances, ensuring robustness across varying scales and complexities. Figure shows how the final output is fused using priority weighting.  This dual-architecture strategy effectively addresses class imbalance and detail loss challenges, resulting in enhanced segmentation performance, particularly for small and critical surgical objects.

\subsubsection{Priority-based Conditional Fusion Strategy}
we fused outputs from the two SegFormer model instances using a priority-based conditional fusion strategy via the OR operation. This method ensures that any region identified by either model is included in the final segmentation, balancing the contributions of anatomical and tool-focused segmentation models. The OR operation was selected for its computational simplicity and effectiveness in retaining critical segmentation without introducing significant noise or ambiguity. In practical scenarios, particularly in surgical environments where tools may obscure anatomical structures, overlapping segmentation present a unique challenge. To address this, (Eq.  \ref{eq:priority_fusion})  shows our fusion strategy incorporates a confidence-weighted scheme, where segmentation from the models are prioritized based on their respective confidence scores. Furthermore, a post-processing step applies morphological operations to refine boundaries and resolve ambiguities in overlapping regions. This refinement ensures that the fused segmentation is both accurate and consistent, even in complex scenes with significant tool-anatomy interactions. 
 
\begin{equation}
C(x, y) = 
\begin{cases} 
M_{\text{inst}}(x, y) & \text{if } P_{\text{inst}} > P_{\text{anat}} \text{ or } M_{\text{anat}}(x, y) = \mathbf{0} \\
M_{\text{anat}}(x, y) & \text{otherwise}
\end{cases}
\label{eq:priority_fusion}
\end{equation}

\subsection{Experimental Setup}
 Two GPUs were used to evaluate the model’s performance: a local NVIDIA RTX4090 and a cloud-based NVIDIA V100‑32G. Larger models were trained on the cloud to reduce runtimes. The code was implemented in PyTorch, and hyperparameters were selected empirically. We used the Adam optimizer with weight decay of $10^{-4}$ and a learning rate of $5\times10^{-6}$, enabling the model to learn fine details while avoiding early plateaus. A cyclic learning rate scheduler and a batch size of 4 were employed to help the model escape local minima across 100 training epochs.

To address class imbalance between extensive background regions and smaller, complex instrument areas, we implemented a combined loss function (Eq. \ref{eq3}) that integrates Tversky loss (Eq. \ref{eq1}) with cross-entropy loss (Eq. \ref{eq2}). We applied geometric augmentations—flips, cropping, and rotations—that preserved color distribution and maintained segmentation precision. 

 We set $\alpha=0.7$ and $\beta=0.3$ to penalize false negatives, enhancing the segmentation of delicate structures such as suturing needles and instrument shafts. This configuration enabled consistent improvement in Dice and mIoU scores while avoiding overfitting.
 
    \noindent \textbf{Equation 1:} Tversky Index formula \cite{Teversky}.
    \begin{equation}
        \text{Tversky Index} = \frac{TP}{TP + \alpha \cdot FP + \beta \cdot FN}
        \label{eq1}
        \end{equation}
        \noindent \textbf{Equation 2:} Cross Entropy formula \cite{crossentropy}.
        \begin{equation}
        \text{Cross Entropy} = - \sum_{i=1}^{N} y_i \log(\hat{y}_i)
        \label{eq2}
        \end{equation}
        \noindent \textbf{Equation 3:} Combined loss.
        \begin{equation}
        \text{Combined\_Loss} = \alpha \cdot \text{Tversky\_Loss} + (1 - \alpha) \cdot \text{CE\_Loss}
        \label{eq3}
    \end{equation}

\section{Results and Discussion}

We thoroughly assessed Surg-SegFormer on two publicly available benchmarks for robot-assisted surgery—EndoVis2017 \cite{EndoVis2017} and EndoVis2018 \cite{EndoVis2018} —and compared it with SOTA models. The model demonstrated notable performance on classes with subtle structures. EndoVis2017 focuses on segmenting seven instruments: Bipolar Forceps, Prograsp Forceps, Large Needle Driver, Vessel Sealer, Grasping Retractor, Monopolar Curved Scissors, and Ultrasound Probe. On the other hand, EndoVis2018 is divided into two tasks: Task 1 (Holistic scene segmentation) originally contains 12 labels spanning anatomy and instrument parts; following common practice, we merge the three fine-grained part labels—instrument shaft, wrist, and clasper—into a single Robotic Instrument Part  class, yielding seven labels for per-class analysis: Background Tissue, RI, Kidney Parenchyma, Covered Kidney, Small Intestine (SI), Suturing Needle (SN), and UP. Task 2 (instrument-type segmentation) likewise comprises seven categories—BF, PF, LND, MCS, UP, Suction Instrument (SI), and Clip Applier (CA). Across both datasets and tasks, Surg-SegFormer achieved SOTA performance, with particular increase over SOTA in SN and UP classes. 

\begin{figure}[ht!]
\raggedright
    \includegraphics[width=0.45\textwidth]{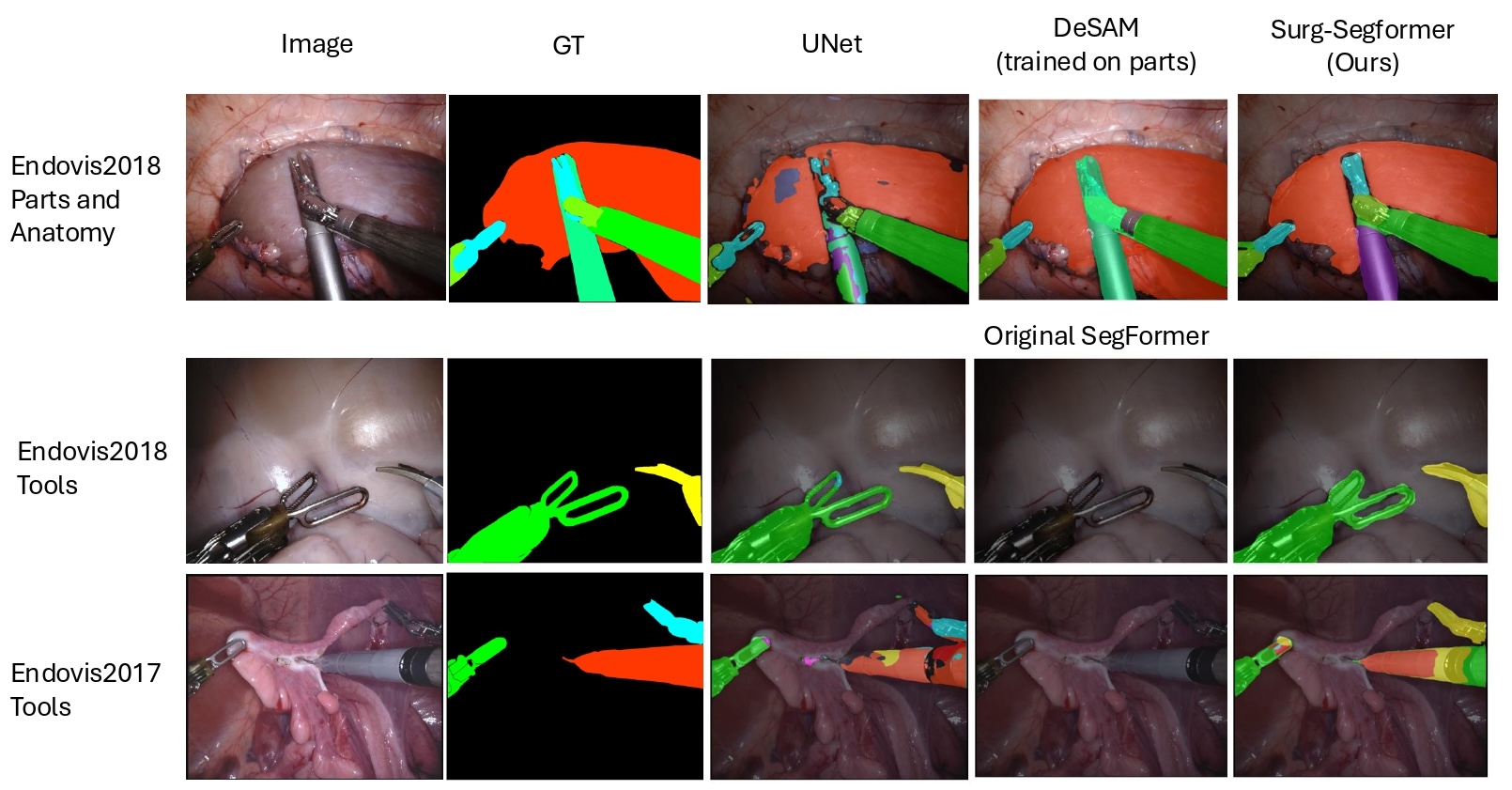}
    \caption{Models' Performance on Different Segmentation Tasks}
    \label{Thumbnails}
\end{figure}

\begin{table}[ht!]
\caption{Models' Overall Performance}
\label{tab:OverallPerformance}
\centering
\renewcommand{\arraystretch}{1.5} 
\setlength{\tabcolsep}{4pt} 
\begin{tabular}{lcccccc}
\hline
\multirow{2}{*}{\textbf{Model}} & \multicolumn{2}{c}{\textbf{Parts 2018}} & \multicolumn{2}{c}{\textbf{Type 2018}} & \multicolumn{2}{c}{\textbf{Type 2017}} \\ \cline{2-7} 
                                & \textbf{mIoU} & \textbf{Dice} & \textbf{mIoU} & \textbf{Dice} & \textbf{mIoU} & \textbf{Dice} \\ \hline
UNet*                          & 0.53        &    0.58         & 0.57        &   0.60          & 0.49        & 0.51            \\ 
SurgicalSAM   \cite{SurgicalSAM}       &      -       &     -        &       \textbf{0.80}     &   -          &    \underline{0.70}         &      -       \\ 
TernausNet  \cite{SurgicalSAM}      & -      &      -       & 0.40        &      -       & 0.13        &    -         \\ 
ISI-Net \cite{SurgicalSAM}        &   -          &    -         & 0.71      &     -        & 0.52       &        -     \\ 
S3Net \cite{SurgicalSAM}     &    -         &   -          & 0.74        &     -        & \textbf{0.72}        &      -       \\ 
MATIS  \cite{SurgicalSAM}         &    -         &     -        & \underline{0.77}        &    -         & 0.63        &    -         \\ 

MedT   \cite{AdaptiveSAM}        & 0.64       &  0.68          &   -          &        -     &    0.29         &       0.31      \\ 
AdaptiveSAM \cite{AdaptiveSAM}                 & \underline{0.65}        &     \underline{0.69}        &    -         &  -           &    \textbf{0.72}         &    \textbf{0.74}        \\ 
SAM-ZS    \cite{AdaptiveSAM}                  & 0.06        &    0.10         &     -        &   -          &    0.03         &     0.06        \\ 
SegFormer* (Single Model)      & 0.57        &    0.59         & 0.46        &      0.47       & 0.41        &      0.42       \\ 
Surg-SegFormer*               & \textbf{0.80}        & \textbf{0.89}     &  0.64           &   \textbf{0.66}          &     0.54        &      \underline{0.56}      \\ 

\hline 
\\
\end{tabular}
\footnotesize{Parts 2018: EndoVis2018 dataset with seven classes (anatomy and tools parts). Type 2018: EndoVis2018 dataset tools type only. Type 2017: EndoVis2017 dataset tools type only. The * means that we trained the models from our side and reported the results. The rest of the results were taken from the models' papers.}
\end{table}

%------------------------------------------------------------------
\subsection{EndoVis2017}
\label{sec:endo17}
On the seven-instrument \textit{EndoVis2017} benchmark (see Table~\ref{tab:OverallPerformance}), Surg-SegFormer achieved an overall \textbf{mIoU of 0.54} and \textbf{Dice of 0.56}. Recent prompt-driven models such as AdaptiveSAM reported higher mIoU (0.72); Surg-SegFormer clearly outperforms the canonical U-Net (0.49/0.51), the re-trained SegFormer backbone (0.41/0.42), and the task-specific ISI-Net (0.52). Class-wise inspection underscores the model’s strength on fine tools. Results in table \ref{tab:EndoVis2017}  shows the model's best IoUs on three classes: \textit{Ultrasound Probe} (\textbf{0.87}) and \textit{Monopolar Curved Scissors} (\textbf{0.69}). Lower scores for \textit{Bipolar Forceps} (0.24) and \textit{Prograsp Forceps} (0.16) likely stem from limited visual diversity and inter-class ambiguity among graspers, yet overall the method delivers balanced segmentation without task-specific tuning.

\begin{table}[ht!]
\caption{mIoU Values on EndoVis2017 - Tools Type}
\label{tab:EndoVis2017}
\centering % IEEE recommends this instead of \begin{center}
\renewcommand{\arraystretch}{1.2} % Adjust row height for better spacing
\setlength{\tabcolsep}{4pt} 

\begin{tabular}{lccccccc}
%\begin{tabular}{p{1cm}m{0.5cm}m{0.5cm}m{0.5cm}m{0.5cm}m{0.5cm}m{0.5cm}m{0.5cm}}
\hline
Model          & BF  & PF  & LND & VS & GR  & MCS & UP  \\ \hline
UNet*           & 0.27  &  0.20  & 0.39 & 0.35  &  \underline{0.41} &  \underline{0.65}   &  0.65\\
%SurgicalSAM    &   0.68  &  0.52   &  0.76   & 0.68    & 0.58    & 0.87    & 0.61    \\
TernausNet \cite{SurgicalSAM}  &  0.13   &  0.12   & 0.21&  0.06   &  0.01   &  0.01   & 0.17    \\
ISI-Net \cite{SurgicalSAM}   &   0.39  & 0.39    & 0.50    & 0.27    & 0.02    & 0.29    &  0.13   \\
S3Net   \cite{SurgicalSAM}    &  \textbf{0.75}   &  \textbf{0.54}   &  \textbf{0.62}   & \underline{0.36}    & 0.27   &  0.43   & 0.28    \\
MATIS \cite{SurgicalSAM}       &  \underline{0.66}   &  \underline{0.51}   & \underline{0.52}    &  0.33  &  0.16   &  0.19   &  0.24   \\
SegFormer*          &   0.00  &  0.0 & 0.003   & \textbf{0.45}  &  \textbf{0.47}   & 0.49 &  \textbf{0.97}   \\
\textbf{Surg-SegFormer*} & 0.24  &  0.16   &  0.47   & \textbf{0.45}    & \textbf{0.47}    &  \textbf{0.69}   &  \underline{0.87}   \\ \hline
\end{tabular}

\vspace{2mm} % Add vertical spacing before the legend

\footnotesize{BF: Bipolar Forceps, PF: Prograsp Forceps, LND: Large Needle Driver, VS: Vessel Sealer, GR: Grasping Retractor, MCS: Monopolar Curved Scissors, UP: Ultrasound Probe}
\end{table}

\subsection{EndoVis2018}
\label{sec:endo18}
For \textit{EndoVis2018}—which comprises Task 1 (scene segmentation with seven merged anatomy + part labels) and Task 2 (instrument-type segmentation)—Surg-SegFormer sets a new state of the art  on Task 1 with \textbf{0.80 mIoU} and \textbf{0.89 Dice}, surpassing MedT (0.64/0.68) and AdaptiveSAM (0.65/0.69) presented in table \ref{tab:OverallPerformance}. These gains are most pronounced in visually subtle classes such as \textit{Kidney Parenchyma} and \textit{Suturing Needle}, for which several baselines fail entirely. On Task 2, the model remains competitive at \textbf{0.64 mIoU / 0.66 Dice}, exceeding re-trained SegFormer (0.46/0.47) and U-Net (0.57/0.60), while trailing SurgicalSAM (0.80 mIoU) and MATIS (0.77 mIoU). Taken together, the results confirm that Surg-SegFormer generalises robustly across anatomy-rich scenes and pure instrument settings, with its strongest margins emerging when complex tissue appearance and fine-grained part boundaries must be resolved. A closer look at the class-wise results in Table \ref{tab:EndoVis2018} reveals why Surg-SegFormer attains the highest overall scores. The model attains the best mIoU in 8 / 10 labels, with particularly large margins on \textit{Robotic Instrument Part} (\textbf{0.70} vs.\ 0.60 for the next best), \textit{Covered Kidney} (0.64 vs.\ 0.44 for MedT), and the extremely thin \textit{Suturing Needle}, where it reaches an almost perfect \textbf{0.98}. It also pushes the upper bound on challenging deformable tissue, achieving \textbf{0.45} on \textit{Kidney Parenchyma}—surpassing every baseline by at least 2 pp and doubling the score of SegFormer. Conversely, its lowest relative advantage is on \textit{Small Intestine} (0.72), where MedT attains 0.78; visual inspection suggests that specular highlights and rapid peristaltic motion still challenge the current attention granularity.

\begin{table}[ht!]
\caption{mIoU Values on EndoVis2018 - Anatomy and Parts}
\label{tab:EndoVis2018}
\centering
\renewcommand{\arraystretch}{1.8} % Slightly tighter row spacing
\setlength{\tabcolsep}{5pt} % Reduce column spacing

% Resize to fit in two columns
\resizebox{\columnwidth}{!}{%
\begin{tabular}{lcccccccccc}
\hline
\textbf{Model}      & \textbf{BT}  & \textbf{RI}  & \textbf{KP} & \textbf{CK} & \textbf{SI}  & \textbf{ST}  & \textbf{SN} & \textbf{Clamps} & \textbf{Suction} & \textbf{UP}   \\ \hline
UNet*              & 0.66        & 0.52        & \underline{0.43}      & 0.07     & 0.18       & 0.38            &\underline{0.91}      &0.76             & 0.70               & 0.72          \\
TransUNet  \cite{AdaptiveSAM}       & 0.59        & \underline{0.60}         & 0.32        & 0.33       & 0.66        & 0.01             & 0.04       & 1               & 0.64               & 0.61             \\
MedT       \cite{AdaptiveSAM}      & 0.40      & 0.54  & 0.16 & \underline{0.44}        & \textbf{0.78}         & 0.82            & 0.90      & \underline{0.96}               & 0.61               & 0.84            \\
AdaptiveSAM \cite{AdaptiveSAM} &0.51 &0.57  & 0.21        & 0.33        & 0.55         & \textbf{0.88}            & \underline{0.91}     & 0.82              & \textbf{0.86}                & \underline{0.85}           \\
SAM-ZS    \cite{AdaptiveSAM}           &0.26     & 0.10 & 0.16 & 0.04         & 0             & 0       & 0              & 0 &0.02              & 0.01             \\
SegFormer*           & \underline{0.87}        & 0.45        & 0.12        & 0.32        & 0.27 & 0.15 & 0.28 & 0.67           & 0.55               & 0.10             \\
\textbf{Surg-SegFormer*}  & \textbf{0.89 }      & \textbf{0.70 }       & \textbf{0.45}         & \textbf{0.64} & \underline{0.72} & \underline{0.86}    & \textbf{0.98}  & \textbf{0.99}           & \underline{0.79}               & \textbf{0.95}         \\ \hline
\end{tabular}
}
 \vspace{2mm}

\footnotesize{BT: Background Tissue, RI: Robotic Instrument Part, KP: Kidney Parenchyma, CK: Covered Kidney, SI: Small Intestine, ST: Suturing Thread, SN: Suturing Needle, UP: Ultrasound Probe}
\end{table}

For the instrument-type task Table \ref{tab:EndoVis2018-type} illustrates the strong performance of Surg-SegFormer, which remains among the top three in five of seven tools, leading on \textit{Suction Instrument} (0.83) and nearly matching the best on \textit{Clip Applier} (0.93). Lower IoUs for \textit{Prograsp Forceps} (0.13) and \textit{Large Needle Driver} (0.09) echo trends already seen in EndoVis2017 and can be traced to visually similar end-effectors and a paucity of examples in the training split. These fine-grained insights confirm that Surg-SegFormer’s hybrid scale-aware design excels when subtle structural cues differentiate classes, while leaving room for future work on grasper-type instruments with high intra-class variance.

%------------------------------------------------------------------

\begin{table}[ht!]
\caption{mIoU Values on EndoVis2018 - Tools Type}
\label{tab:EndoVis2018-type}
\centering % IEEE recommends this instead of \begin{center}
\renewcommand{\arraystretch}{1.2} % Adjust row height for better spacing
\setlength{\tabcolsep}{4pt} 
\begin{tabular}{lccccccc}
\hline
Model          & BF  & PF  & LND & MCS & UP  & SI  & CA  \\ \hline
UNet*           & 0.64 & 0.12 & 0.12 & 0.66  & 0.54 &  0.73 &  0.8  \\
%SurgicalSAM    & 0.84 & 0.66 & 0.59 & 0.89 & 0.21 & 0.54  & 0.40 \\
TernausNet \cite{ternaus}   & 0.44 & 0.05 & 0.00 & 0.50 & 0.00 & 0.00 & 0.00 \\
ISI-Net  \cite{SurgicalSAM}   & 0.74 & \underline{0.49} & \underline{0.31} & 0.88 & 0.02 & 0.38 & 0.00\\
S3Net  \cite{SurgicalSAM}   & \underline{0.77} & \textbf{0.50} & 0.20 & \underline{0.92} & 0.07 & 0.51 & 0.00\\
MATIS     \cite{SurgicalSAM}   & \textbf{0.83} & 0.39 & \textbf{0.40} & \textbf{0.93} & 0.16 & 0.64 & 0.04\\
SegFormer*          &   0.04  &  0.05 & 0.08   & 0.20  & \textbf{0.76} & \underline{0.80}&  \textbf{0.95}   \\
\textbf{Surg-SegFormer*} &  0.67   &  0.13   &  0.086   &  0.82  &   \underline{ 0.70}  & \textbf{0.83}  &  \underline{0.93}   \\ \hline
\end{tabular}

\vspace{2mm} % Add vertical spacing before the legend

\footnotesize{BF: Bipolar Forceps, PF: Prograsp Forceps, LND: Large Needle Driver, MCS: Monopolar Curved Scissors, UP: Ultrasound Probe, SI: Suction Instrument, CA: Clip Applier.}
\end{table}

\subsection{Performance Analysis}
\label{sec:benchmarks}

Table \ref{tab:OverallPerformance} reveals that several baseline architectures excel on a single benchmark yet underperform on the other.  S3Net and AdaptiveSAM, for instance, lead the instrument–only EndoVis2017 task (mIoU 0.72) but fall to 0.74 and 0.65, respectively, when anatomy and part labels are introduced in EndoVis2018.  A complementary pattern appears for MATIS, which ranks near the top for EndoVis2018 instrument–type segmentation (0.77 mIoU) yet drops to 0.63 on EndoVis2017. 

Surg-SegFormer presents a more balanced profile.  Although its 0.54 mIoU on EndoVis2017 trails the prompt-tuned leaders by roughly eighteen percentage points, it remains well ahead of classical U-Net (0.49) and the re-trained SegFormer backbone (0.41).  The same architecture rises to the top of EndoVis2018 Task 1 with 0.80 mIoU and 0.89 Dice, outperforming the strongest transformer-based baseline, MedT, by sixteen percentage points. Qualitative examples in Fig. \ref{Thumbnails} confirm the numerical trend: in scenes with overlapping tools and ambiguous tissues, baseline outputs either smooth away fine structures or miss entire parts, whereas Surg-SegFormer retains complete masks and sharp boundaries.  

The consistency across the two datasets is attributed to three design elements: a dual-branch encoder that specialises separately in tissue and metallic cues; a priority-weighted fusion rule that reduces false negatives in crowded frames; and a hybrid Tversky–cross-entropy loss that counteracts background dominance while preserving sub-pixel detail. 

\subsection{Ablation Study}

Our ablation study was conducted to evaluate the impact of various configurations on the performance of the Surg-SegFormer model. The model went through different configurations for different aspects, such as  loss functions, and the data fusion operation of the dual model output.

\textit{Loss Functions } We chose the model's loss function through examining various single-loss-function methodologies versus composite loss methodologies. The combined loss function in our model integrates Tversky Loss and Cross-Entropy Loss, addressing the inherent class imbalance present in surgical datasets dominated by background pixels. Tversky loss excels at managing class imbalance and highlighting the boundaries of segmented objects, whereas multi-class cross-entropy is superior in achieving overall classification accuracy across multiple classes. 
The synergistic Tversky loss and multi-class cross-entropy use their strengths to improve training. This approach effectively penalizes false negatives, particularly for small and intricate objects like suturing needles and tool-tips, ensuring better delineation against complex surgical backgrounds. 

First, we tested the single-loss functions approach—Tversky loss function and multi-class cross-entropy loss,   then compared it with the combination of the two. To achieve the best balance between class imbalance and classification accuracy, the parameters $\alpha$ and  $\beta$ are optimized through testing. We found that the configuration, with $\alpha = 0.7$ and $\beta = 0.3$, prioritizes the penalization of false negatives to enhance segmentation accuracy for challenging regions. As seen in Table \ref{tab:loss_comparison}, our combined strategy had the highest mIoU of 85.7\% and Dice coefficient of 89.21\%, outperforming the single-loss approaches, and demonstrating that combining Tversky loss with multi-class cross-entropy effectively optimizes segmentation accuracy and addresses class imbalance. 

\begin{table}[ht!]
\centering
\scalebox{1.0}{ % Adjust the scale factor to control font size
\begin{tabular}{p{3cm}p{1cm}p{1cm}}
\hline
\textbf{Loss Function} & \textbf{mIoU} & \textbf{Dice} \\ 
\hline
Tversky & 64.79 & 65.54 \\ 
Cross-entropy & 74.34 & 76.18 \\ 
Combined loss &\textbf{ 85.70}& \textbf{89.21 }\\ 
\hline
\end{tabular}
}
\caption{Loss Functions Comparisons}
\label{tab:loss_comparison}
\end{table}

 Future refinements could explore dynamic loss weighting schemes, where weights adjust adaptively based on the proportion of each class within a frame. Additionally, focal Tversky Loss could be incorporated to down-weight well-classified regions while emphasizing hard-to-segment examples. These enhancements would further improve segmentation robustness in highly imbalanced datasets, paving the way for more accurate and generalizable models.

\section{CONCLUSION}

In this work, we presented Surg-SegFormer, a unified and lightweight transformer-based architecture tailored for surgical scene understanding. Unlike many existing models that specialize in either anatomical or tool segmentation, Surg-SegFormer addresses both tasks simultaneously, demonstrating strong performance in multi-class and single-class surgical segmentation. Through extensive experiments on the EndoVis2017 and EndoVis2018 datasets, our model consistently outperformed classical and recent SoTA approaches, including prompt-based methods, particularly in anatomically complex or visually challenging scenes. This suggests that Surg-SegFormer can serve as a robust backbone for real-time, intraoperative surgical assistance systems, providing precise segmentation of both instruments and critical anatomy.

The high segmentation accuracy—achieved without reliance on handcrafted prompts, large models, or heavy post-processing—emphasizes the efficiency and scalability of our approach. The incorporation of a hybrid loss function (Tversky + Cross-Entropy) proved particularly effective in handling class imbalance, contributing to more stable training and better performance across underrepresented categories.

\addtolength{\textheight}{-12cm}   % This command serves to balance the column lengths
                                  % on the last page of the document manually. It shortens
                                  % the textheight of the last page by a suitable amount.
                                  % This command does not take effect until the next page
                                  % so it should come on the page before the last. Make
                                  % sure that you do not shorten the textheight too much.

%%%%%%%%%%%%%%%%%%%%%%%%%%%%%%%%%%%%%%%%%%%%%%%%%%%%%%%%%%%%%%%%%%%%%%%%%%%%%%%%

%%%%%%%%%%%%%%%%%%%%%%%%%%%%%%%%%%%%%%%%%%%%%%%%%%%%%%%%%%%%%%%%%%%%%%%%%%%%%%%%

%%%%%%%%%%%%%%%%%%%%%%%%%%%%%%%%%%%%%%%%%%%%%%%%%%%%%%%%%%%%%%%%%%%%%%%%%%%%%%%%

\section*{ACKNOWLEDGMENT}

The authors would like to acknowledge the support of the Surgical Research Section at Hamad Medical Corporation for the conduct of this research.

Research reported in this publication was supported by the Qatar Research Development and Innovation Council (QRDI) grant number ARG01-0522-230266. The content reported through this research is solely the responsibility of the authors and does not necessarily represent the official views of Qatar Research Development and Innovation Council.

\end{document}